# The Prediction of Intrinsic Wettability without Young's Equation

Haohao Gu, Hao Wang

From College of Engineering, Peking University

wanghpku@pku.edu.cn

1. Introduction

Wetting phenomena widely exist in both production and fundamental scientific researches. With a solid substrate exposed to open environment, it is inevitably covered by liquid film or droplets. As Danniel Bonn et al. reviewed in 2009, wetting phenomena are a playground where chemistry, physics and engineering intersect[1]. Ever since Young[2] proposed his most famous equation on wetting phenomena, although new findings and knowledge keep emerging with the development of observation and research techniques[4-18], Young's equation keep dominating this area for over 200 years. Due to the complexity rooted in the interlaced nature of wetting, a thorough and fundamental understanding of liquid wetting on solid surface with complex topology and chemical composition is still lacking. Hao Wang reviewed existing research works concerning the static-to-dynamic process of liquid wetting and spreading, he also suggested the intrinsic drawback of Young's equation – the lack of predictive ability due to the unrealistic assumptions and dependency on macroscopic properties, indicated that the relative relation between interaction among liquid molecules (cohesion) and solid-liquid particle interactions (adhesion) determine the intrinsic static contact angel with specific surface conditions (topology, defects, heterogeneity and etc.) bringing complexity, such conclusion outlined a unified framework that includes number of major topics in the interfacial area such as: contact angle hysteresis, wetting, surface adsorption, disjoining pressure[3]. Therefore, to have a thorough understanding of wetting phenomena and to predict the wetting ability of a specific combination of liquid, solid molecules and surface conditions, the very first step is to investigate the relationship between intrinsic wetting ability and inter-particle actions while neglecting all the surface properties, on the base of the understanding of ideal planar surface wetting ability by introducing specific surface characteristics finally provide a solution to predict the contact angle of realistic surface and various liquids.

To construct an ideal solid surface to ignore any specific surface characteristics means the solid phase should be homogeneous to ensure only one solid-fluid interaction is involved, and surface roughness is another factor that controls wettability. Though the role played by solid roughness has long been accepted in wetting phenomena[4-7], but the definition of absolute planar surface still under debate. Researchers used to take root-mean-square roughness 100 nm as a threshold to distinguish absolute smooth surface which can eliminate any contact angle hysteresis[10]. Trammer et al.[8], Lam et al.[9]'s experiments proved that significant hysteresis still exists even on solid surfaces with root-mean-square (RMS) roughness below 20 nm (e.g. silicon wafer, polymer, polyamide). These results suggest that even far below the conventional roughness threshold, the surface topography of a homogenous solid surface, the rough elements of ten nanometers can still exert influence on intrinsic wettability that cannot be ignored. Therefore, even surfaces smooth like mica or silicon wafer cannot fulfill the goal to eliminate any possible effect led by specific surface characteristics and investigate the role of micro inter-particle-actions solely. Beforementioned facts suggest the infeasibility of experimentally investigate the relationship between molecular interactions and intrinsic contact angles. Molecular dynamic simulation becomes an appropriate solution for such purpose.

The interactions between a pair of micro particles are described by the conventional form of Lennard-Jones(LJ) potential (12-6), and by integrating the potential energy between a half-infinite space with LJ particles uniformly distributed and a isolate LJ particle can get the 9-3 form of Lennar-Jones potential, which describes the interaction between a liquid molecule and a mathematically ideal planar surface that ignore any specific surface characteristics[11].

$$\varphi_{ls}(x) = \frac{2}{3}\pi\varepsilon_{ls}\rho_s\sigma_{ls}^3[\frac{2}{15}\left(\frac{\sigma_{ls}}{x}\right)^9 - \left(\frac{\sigma_{ls}}{x}\right)^3] \quad (1)$$

Equation (1) is the Lennard-Jones 9-3 potential, where $\varepsilon_{ls}$ and $\sigma_{ls}$ are the energy and distance parameters in Lennard-Jones 12-6 potential respectively, x is the distance between the isolating molecule and the half space, the subscripts l and s represent different kind of particles. In 1981, Soren Toxvared et al. first adopted LJ potential to investigate the solid-fluid interface consists of (111) crystal plane and LJ fluids, which is the very first application of MD simulation in solid-fluid interaction area. With the progress in computational power, researchers could afford greater calculation cost to simulate larger more complicated systems and get a better understanding of wetting phenomena. A few researchers introduced the Lennard-Jones 9-3 potential and investigated wetting phenomena on ideal planar surface, like Sharaz et al.[12], T. Ingebrigtsen and S. Toxvaerd et al.[13]. While many researchers turned to realistic surface which have specific surface defects. D. Toghraie Semiromi et al.[14] focused on a sessile drop of argon molecules spreads over a platinum surface with lattice structure (with no heterogeneity or surface defects) and reported the temperature dependency of argon droplet spreading. Bo Shi and Vijay K. Dhir et al.[15] compared contact angles of LJ fluids and water and other wetting properties of nano-droplets. Other researchers also introduced heterogeneity to solid material or different kinds of surface defects. Mathias Lundgrend et al. investigated the relationship between contact angle and solid surface patterns[16]. Sandeep Pal et al.[17], Janne T. Hirvi et al.[18] proposed models to describe the dependency of contact angle of nanodroplet on nano-grooved surfaces. In 2009, Xin Yong[19] et al. compared the simulation results of a sessile mercury droplet on cooper surface with different groove patterns, from merely lattice order to nanogroove patterns, they also investigated the effect exerted by energy parameters to wettability. In 2019, Qiao Liu et al.[19] adopted Lennar-Jones 9-3 potential when investigated the role played by roughness in contact angle hysteresis, and reported that to achieve same contact angle the solid-fluid molecular energy parameter of LJ 9-3 potential has to be significantly larger than the one in LJ 12-6 potential even with same molecules.

Before mentioned researches have covered the entire wetting phenomena of nano-droplet on solid surfaces from ideal planar to realistic surface with lattice structures and defects, however, none of these answered the question that what lead to the significant difference between contact angles on ideal smooth surface and on surface with real lattice structure. This work employed two surfaces consist of identical solid atoms, one with mathematically ideal smooth surface one with lattice structure in molecular dynamic simulation to distinguish the factors that lead to the wetting differences on ideal and realistic surfaces. Based on the comparison results, the currently research also proposed a solution to use the competition of cohesion (liquid-liquid) and adhesion (liquid-solid) energy to predict wettability of specific combination of fluid and solid particles.

2.simulation and experiment methods

Molecular dynamics simulation was employed to investigate droplet spreading on solid surfaces with two structures – ideal planar and lattice structure. A sessile drop consists of 12500 argon atoms is setted on two surfaces with diameter up to 30 nm. The size of the simulation boxes depends on the lattice constant chosen, when platinum was chosen the dimensions were (x-length, y-length, z-length) = (40, 35, 3)nm, the motion time integration was carried out using velocity Verlet algorithm[20], time step was set at 3.0 fs and the temperature for argon system was chosen to be 82 k while the macro properties of argon during simulation agreed with experimental results[21]. The simulation was first carried out in the microcanonical ensemble (NVE) for 100000 steps to relaxation and minimize potential energy, during which the temperature was controlled by velocity rescaling method, then kept in the canonical ensemble (NVT) with Nose-Hoover thermostat[22] adopted to maintain constant temperature and transport properties of the liquid phase at the same time. LAMMPS software package which has been widely applied to wetting researches[13,19] was adopted.

The before mentioned Lennard-Jones 9/3 wall was adopted as absolutely planar surface to investigate the intrinsic wettability while ignore any specific surface properties. As shown in equation 1, the interaction was the integration of a single liquid particle with a half-infinite space filled with LJ particles uniformly distributed.

Surfaces consist of platinum atoms with lattice structure served as a representation of realistic solid surfaces as comparison. This surface was atomically smooth single FCC crystal with lattice constant 0.392 nm. The interaction between this solid surface and liquid particles was described using the classical Lennard-Jones potential with 12/6 form i.e.

$$\varphi_{ls}(x) = 4\varepsilon_{ls}[\left(\frac{\sigma_{ls}}{x}\right)^{12} - \left(\frac{\sigma_{ls}}{x}\right)^{6}] \tag{2}$$

To acquire accurate results of contact angles and detect the fine structure of droplet interior and phase interfaces, the simulation space was divided into 2-dimension grid with length steps in x, y direction set to 0.1 Å after a thorough relaxation of the system, then continued the simulation and calculated the temporal average of liquid particles' number density of each grid unit in 10000 steps, these results served as data source for further processing and figure constructing. As shown in figure 1, (a) and (b) are the number density distributions of the droplets with identical contact angle on absolutely smooth surface and lattice structure surface with identical contact angle respectively.

For a better understand of the factors that lead to the contact angle difference, two kind of absolutely smooth surfaces were adopted in presented work, compared with lattice structure surface one showed same contact angle when exposed to argon droplet i.e. surface A, another had identical liquid-solid interacting energy parameter and surface C has lattice structure for comparison.

Table 1   Simulation parameters and contact angles

| surface | $\varepsilon_{ls}/kj\cdot mol^{-1}$ | $\varepsilon_{ls}^*/kj\cdot mol^{-1}$ | $\sigma_{ls}/\text{Å}$ | CA/° |
|---------|-------------------------------------|---------------------------------------|------------------------|------|
| A       | 0.1734                              | 0.613                                 | 2.94                   | 61.5 |
| B       | 0.1346                              | 0.4757                                | 2.94                   | 86.6 |
| C       | 0.1346                              | None                                  | 2.94                   | 86.6 |

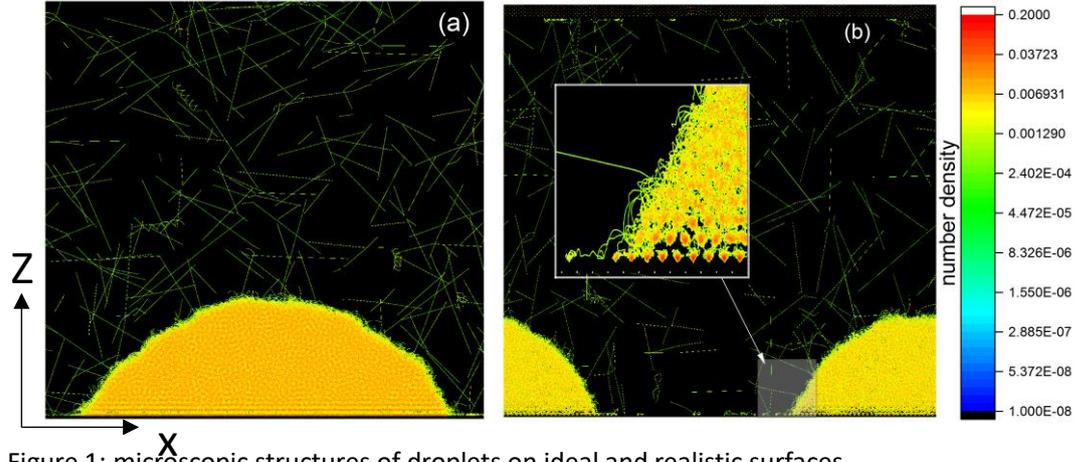

Figure 1: microscopic structures of droplets on ideal and realistic surfaces

3. results and discussion

3.1 intrinsic contact angle prediction

To predict the contact angles, the competition between cohesion (liquid-liquid) and adhesion (liquid-solid) force inside this area should be quantitatively described. Since the cut-off length of molecular dynamics simulation usually chosen to be 3 to 3.5 times of the distance parameter, and based on the layer separation data in last section, 3 layers of liquid particles were chosen to determine the cohesion and adhesion forces. The interaction energy between 3 adjacent liquid layers and solid particles (for absolutely smooth surface is the entire half space) represents the adhesion force. The interaction energy among these liquid layers and inner-layer represents the cohesion force. The competition between these two interactions is represented by the ratio of solid-liquid interaction energy over inner-liquid energy.

a. Solid fluid interaction

For continuous liquid layer on absolutely smooth surface, the continuous form is adopted and result can be easily integrated, which can be expressed as follow. By integrating the LJ 9/3 form, which express the interaction energy between a single particle with a half space filled with LJ particles, to the planar consists of liquid particles can get the energy between a liquid layer and the solid surface.

$$W_{LS} = \frac{3}{4}\frac{R^2 \sigma_{ls}}{\sigma_{ll}^3} \varepsilon_{ls}^* \left[\frac{2}{15}\left(\frac{\sigma_{ls}}{x}\right)^9 - \left(\frac{\sigma_{ls}}{x}\right)^3\right] \quad (2)$$

Where, R is the radius of the liquid layer, $\varepsilon$ and $\sigma$ are energy and distance parameters in LJ 12/6 potential, $\varepsilon_{ls}^*$ is the energy parameter for LJ 9/3 potential, x is the distance between liquid layer and solid surface.

b. Inter layer energy for liquid

The cohesion force is divided into the inner layer part and inter layer part. For absolutely smooth surface, firstly the equilibrium liquid particle density is calculated by minimizing the potential energy of an infinite space uniformly filled with liquid particles, secondly integrating the energy between a single particle and a uniformly distributed thin layer and summing for entire layer lead to the result.

$$W_{LL} = \frac{9}{4} \frac{R^2 \sigma_{ls}^2}{\sigma_{ll}^4} \varepsilon_{ll} [\frac{2}{5}(\frac{\sigma_{ll}}{x})^{10} - (\frac{\sigma_{ll}}{x})^4] \qquad (3)$$

c. Inner layer energy for liquid

The calculation is very similar to the previous part, by integrating the energy between a single particle with a layer with uniformly distributed particles and summing for the whole layer give the result, need to be cautioned is that the interaction here is inside the layer so for each particle would be calculated twice so the result needed to time 0.5.

$$W_{in} = \frac{9}{32} \frac{\pi R^2 \sigma_{ls}^2}{\sigma_{ll}^4} \varepsilon_{ls}^* [\frac{2}{5}(\frac{\sigma_{ll}}{r_0})^{10} - (\frac{\sigma_{ll}}{r_0})^4] \qquad (4)$$

Where $r_0 = \frac{2\sqrt{3}}{3} \frac{\sigma_{ll}^{1.5}}{\sigma_{ls}^{0.5}}$, is the equilibrium distribution distance

Based on eq. 2-4, the interaction energy of adhesion and cohesion force can be expressed, so the energy ration, $\frac{W_{LS}}{W_{LL}+W_{in}}$, represents the competition between them. By changing the MD simulation parameters employed, the particles' properties involved would change consequently and the wettability of the system would vary. And by feeding the simulation parameters into eq. 2-4 can calculate a value of energy ratio corresponding to the contact angle simulated by the specific combination of different parameters. In current work, 38 different combinations of MD parameters and the corresponding contact angles were given based on the aforementioned simulation method, and the results showed a strong linear relationship between energy ratio and contact angles simulated, can be expressed as:

Contact angle $= -179.72 *$ Energy ratio $+ 180.423$ \qquad (5)

3.2 droplet structure differences

To further investigate the factor lead to different wettability of absolutely smooth surface and lattice structure surface and modify the contact angle equation to adapt for lattice structure surfaces, the fine structures of the liquid drops consist of identical particles should be further investigated. As shown in Fig. 1, the obvious layered separated structures exist on both surfaces, but in Fig. 1(b), inner layer periodic structure is quite apparently in several liquid particle layers adjacent to solid surface while on absolutely smooth surface liquid particle seems to obey uniform distribution inside layer. To investigate the difference in the wettability of surfaces, both inner and inter layer periodic structures need to be discussed.

3.2.1 inter layer periodic structure

By averaging liquid particles number density in grid units with same vertical ordinate can get a density distribution curve on the top of surfaces A, B and C. As shown in Fig. 2, each local maximum value of these density curves represent a separated layer structure in liquid drop, and the local density value of each peak obeys exponential decay function. Part (a), (b), (c) are results based on droplets set on surfaces A, B and C.

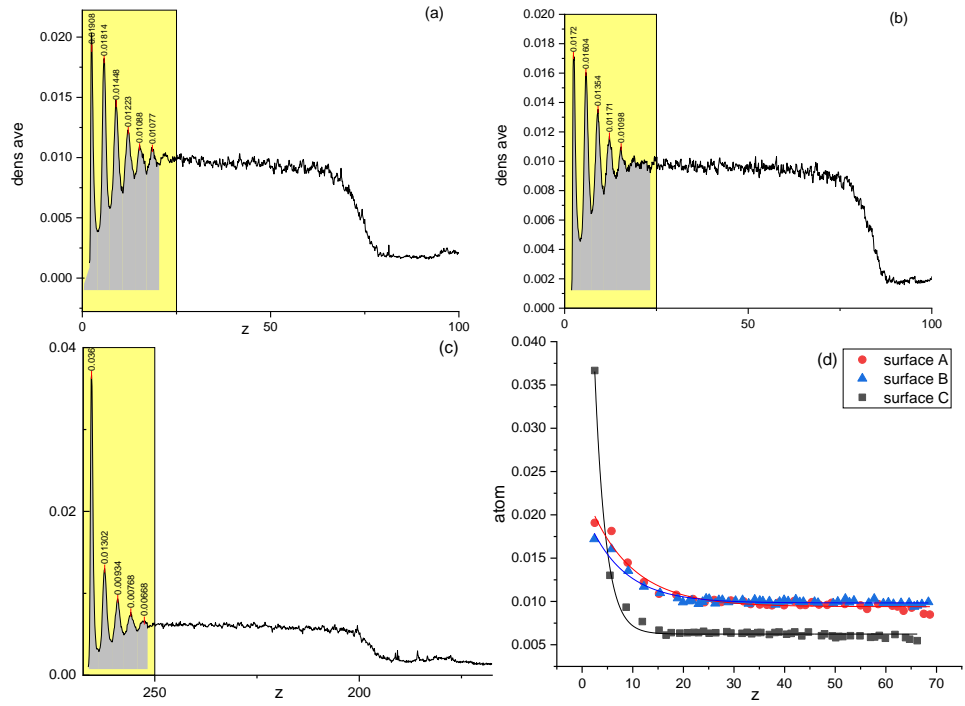

Figure 2: vertical density profile on surface A, B, C and the exponential decay tendency of average density v.s. vertical distance.

The local maximum value of number density represents the exact position where liquid particles most likely to exist, thus can be taken as the accurate vertical coordinate of each liquid particle layer; and the adjacent two local minimum values of the density profile determine how thick the specific liquid layer is. Based on the beforementioned method, the spatial distribution information of the several nearest neighbor liquid layers of solid surfaces can be extracted, as listed out in Table 2. Relative peak height is defined as the ratio of local density maximum value over local minima.

Table 2 density layer separation data

| Surface A | | | | | | |
|---|---|---|---|---|---|---|
| valley | density | peak | density | valley | density | Relative height |
| 1.5 | 0.0015 | 2.1 | 0.0203 | 4 | 0.0038 | 7.660377 |
| 4.3 | 0.0038 | 5.8 | 0.01823 | 7.3 | 0.0058 | 3.797917 |
| 7.3 | 0.0058 | 8.9 | 0.01478 | 10.8 | 0.00705 | 2.300389 |
| 10.8 | 0.00705 | 12.1 | 0.0123 | 13.8 | 0.00825 | 1.607843 |
| 13.8 | 0.00825 | 15.2 | 0.01101 | 17.1 | 0.00862 | 1.305276 |
| 17.1 | 0.00862 | 18.4 | 0.0108 | 19.8 | 0.00926 | 1.208054 |

Surface B

| | | | | | | |
|---|---|---|---|---|---|---|
| 1.9 | 0.00122 | 2.4 | 0.0172 | 4.2 | 0.00456 | 5.951557 |
| 4.2 | 0.00456 | 6.2 | 0.016 | 7.2 | 0.00643 | 2.911738 |
| 7.2 | 0.00643 | 9.5 | 0.0135 | 10.7 | 0.00775 | 1.90409 |
| 10.7 | 0.00775 | 13.2 | 0.0117 | 14 | 0.00882 | 1.412191 |
| 14 | 0.00882 | 16.9 | 0.011 | 17.5 | 0.0093 | 1.214128 |
| | | | Surface C | | | |
| 1.5 | 0.001 | 2.1 | 0.03667 | 4.1 | 0.00289 | 18.85347 |
| 4.1 | 0.00289 | 5.5 | 0.013 | 7.3 | 0.00405 | 3.746398 |
| 7.3 | 0.00405 | 8.7 | 0.00934 | 10.2 | 0.00492 | 2.082497 |
| 10.2 | 0.00492 | 11.9 | 0.00767 | 13.3 | 0.00541 | 1.484995 |
| 13.3 | 0.00541 | 15.2 | 0.00668 | 16 | 0.00576 | 1.196061 |

As shown in Fig. 2, all droplets showed similar density profile and exponential decay tendency from a very large value adjacent to solid surfaces to the equilibrium bulk phase value, but with some subtle differences. The results of two absolutely smooth surfaces are almost identical, while the lattice structure surface is different. Firstly, the first two layers of liquid at lattice structure surface have higher number density, the reason is the inner layer periodic structure is very strong in these layers and most particles concentrated in several positions of the layer while other place left empty. Secondly, because of the higher adhesion force led by the higher energy parameter of surface A, the layered structure extends further into the droplet interior, surface B and C both have 5 layers of liquid particles that can be distinguished while surface A have 6 layers. Thirdly, the most important feature is shown in table 2, even though surface A is more affinitive to liquid particles because of the larger energy parameter, the layer positions are very similar to surface C, one with smaller energy parameter but lattice structure, while the liquid layers' vertical position is a bit higher on surface B which has identical adhesion force.

3.2.2 inner layer periodic structure

Based on the layer separation method described in last section, the space above solid surfaces can be easily divided into different layers of liquid for further investigation of inner layer structure. Since most liquid particles distributed inside the range of each layer which has been listed in Table 2( the range of each layer can be defined by the valley coordinates), the density summation of grid units with same x coordinate in the range of each layer can represent the liquid particle number of this layer at the specific x position. Since the distribution form of liquid layers on different surfaces differ significantly, box-whisker plot has been employed to investigate density distribution features. As shown in Fig. 3, (a) is the comparison of bulk phase density distributions of surfaces A, B and C in a layer with the thickness of 3 Å, (b) listed five neighbor layers density distributions of three surfaces. The bulk phase density distributions of surfaces A, B ad C and similar, all show symmetric distribution around the average density; notably, the average density of surface A and C and similar which is a little bit higher than surface B. Part (b) showed the density distribution neat solid surfaces in detail, the differences between two kind of surfaces arise. For surfaces A and B, the near neighbor layers have same distribution features like bulk phases, which is symmetrically distributed among the average value with outlier points on both sides. This suggests that, for droplets on absolute smooth surfaces, despite of the periodic layered structure vertical to the surfaces, the interior structure of each layer has exactly same distribution form like bulk phase – uniform distribution. For droplets on lattice structure surfaces however, the inner layer periodic structure shows exponential decay with the distance to solid surfaces increasing. In the very beginning layers of surface C's droplet, density asymmetrically distributed with all the outliers significantly higher than mean value, and with the distance increases the asymmetry decays rapidly and finally returns to the bulk distribution.

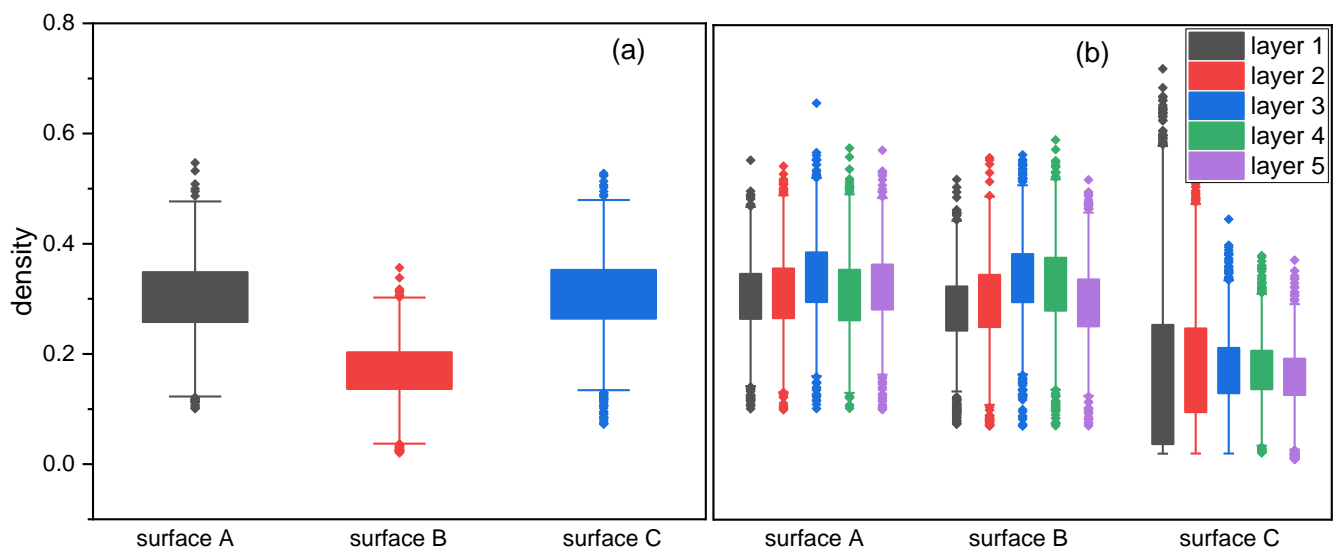

Figure 3: inner layer structures comparison of droplets on surface A, B, C

Based on the layer coordinate data, the density distribution of each layer can be extracted based on the similar method adopted previously. As shown in Fig. 4, inner layer distributions of droplets on lattice and absolutely smooth surfaces are given.

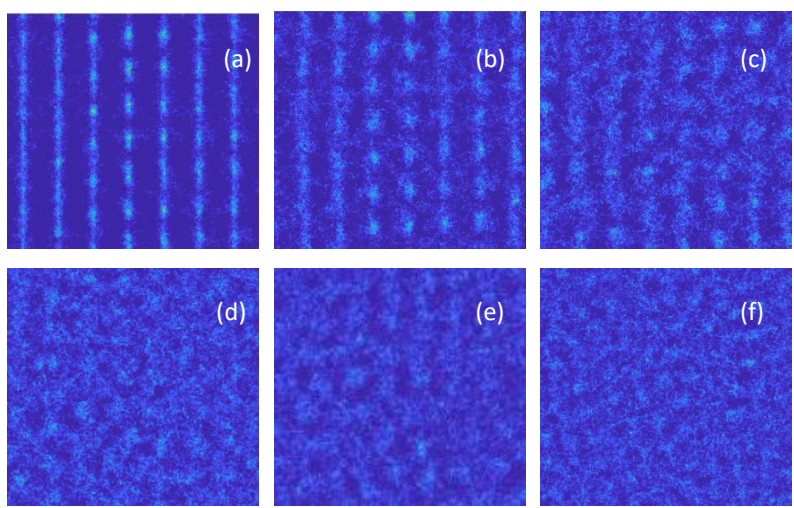

Figure 4: inner layer density distribution of different liquid layers. (a) – (e) are figures of the 1$^{st}$ to the 5$^{th}$ layer of droplet on surface C respectively and (f) is the 1$^{st}$ layer of droplet on surface A for comparison.

The liquid periodic structure rapidly decays with the increasing of solid-liquid distance or the weakening of solid-liquid interaction. Except from the very first layer adjacent to solid surface, most liquid layers do not show completely periodic or uniform distribution but a combination of both distribution features, with stronger solid-liquid interaction and shorter separation distance periodic structure dominates vice versa. Such behavior indicates that the previous energy equation can't be applied to predict the wetting of lattice surfaces and the inner layer distribution features need to be described. The ratio of standard deviation over mean value of inner layer number density therefore is defined as a representation of the degree of deviation from uniform distribution. And for the convenience of calculation, the ratio of the very first layer of a complete wetting droplet is taken as 1 and the ratio of droplet on absolutely smooth surface is taken as 0 with other data linearly mapping into this range, which gives a definition for structural number in range of 0 to 1. Since the periodic structure roots from the interaction exerted by solid while the interaction among liquid itself tend to draw the distribution back to uniform style, the concept of energy ratio can still be adopted to express the competition of the two mechanisms. The results show that the structural number defined previously linearly depends on the single layer energy ratio as shown in Fig.5.

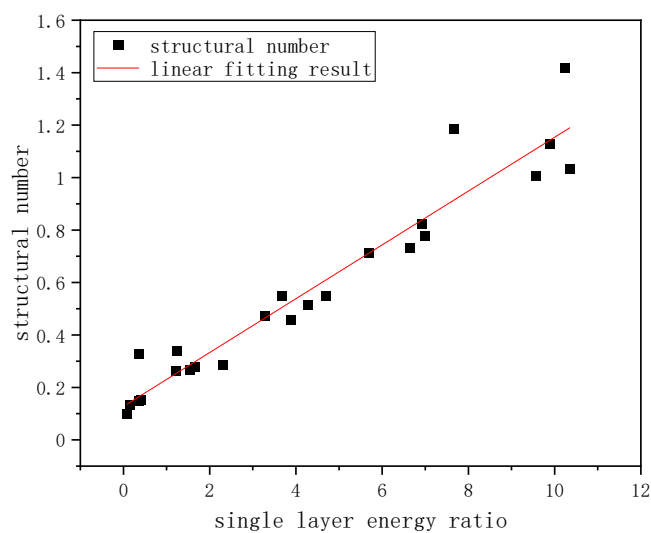

Figure 5: relationship between single layer energy ratio and structural number defined.

The beforementioned features indicated that, the lattice arrangement of solid surface exerts enormous impact on the adjacent layers of liquid molecules and forces them to adopt periodic distribution inside layer. The inner layer structure gradually returns to the uniform distribution of bulk phase, and the differences between droplets on absolutely smooth and lattice surfaces fade out eventually with identical structure in bulk area. Need to be cautioned, the differences due to lattice structure of solid are limited in the few layers of liquid adjacent to solid, therefore the wetting difference should root in this area. As shown in Fig. 6, when the particles that compose the droplets and solid are identical the inter-particle equilibrium distance should be the same which means same $d_0$ and d in Fig. 6. For liquid layers on absolutely smooth surfaces, due to the uniform distribution each part of the layer is equivalent therefore the inter-particle equilibrium distances equal to the inter-layer distances in static droplet. For droplets on lattice surfaces however, the particles in the several adjacent layers are scattered, so the inter-particle distances significantly larger than the final equilibrium distances of liquid layers, therefore, with same liquid and solid particles' combination, the liquid-to-solid distances keep shrinking in the several layers with scattered distribution, and the interaction exerted by solid on liquid is stronger, finally the apparent contact angle of same droplet on realistic surfaces is smaller than the one on absolutely smooth surfaces.

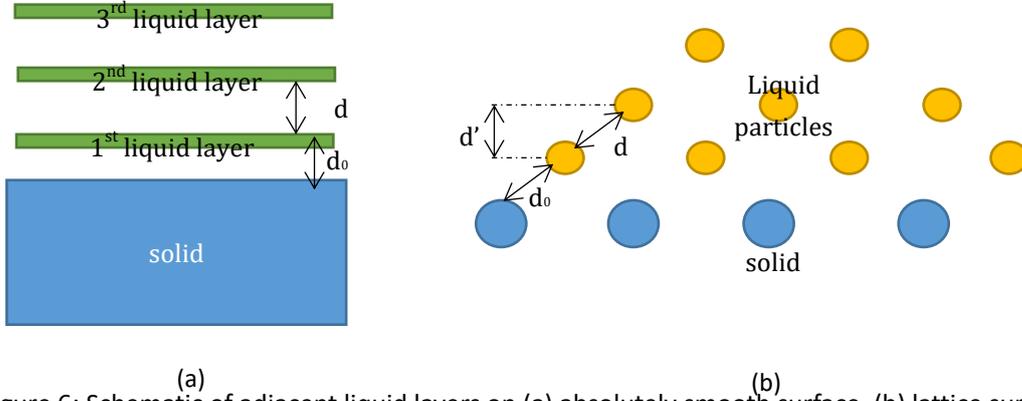

(a) (b)
Figure 6: Schematic of adjacent liquid layers on (a) absolutely smooth surface, (b) lattice surface

3.3 modification for lattice structure surface

As has been discussed in the former section, the structure differences in the few liquid layers adjacent to solid surfaces lead to the significant wetting disparity on absolutely smooth and lattice structure surfaces. Consequently, the fine structural differences are needed to modify the previous energy eq. 2-4 to calculate the adhesion and cohesion forces in droplet on surface with lattice structure. The former expressions of adhesion and cohesion forces are modified, 3 layers of liquid and solid particles within three lattice constants are involved in calculation.

a. Solid fluid interaction

For continuous liquid layer, the interaction energy can be derived by integrating the energy between single particle with the entire solid space.

For scattered particles on surface with lattice structure, the interaction energy can be expressed as the summation of the interaction between a liquid particle and solid atoms within 3 lattice constants, then sum over the entire layer.

$$W_{LS}(lattice) = \frac{R^2 \sigma_{ls}}{\sigma_{ll}^3} \varepsilon_{ls} \left\{ \left(\frac{\sigma_{ls}}{x}\right)^{12} - \left(\frac{\sigma_{ls}}{x}\right)^6 + 4\left[\frac{\sigma_{ls}^{12}}{(x^2+a^2)^6} - \frac{\sigma_{ls}^6}{(x^2+a^2)^3}\right] + 4\left[\frac{\sigma_{ls}^{12}}{(x^2+4a^2)^6} - \frac{\sigma_{ls}^6}{(x^2+4a^2)^3}\right] + 4\left[\frac{\sigma_{ls}^{12}}{\left(\left(x+\frac{\sqrt{2}}{2}a\right)^2+a^2\right)^6} - \frac{\sigma_{ls}^6}{\left(\left(x+\frac{\sqrt{2}}{2}a\right)^2+a^2\right)^3}\right] + 4\left[\frac{\sigma_{ls}^{12}}{\left(\left(x+\frac{\sqrt{2}}{2}a\right)^2+a^2\right)^6} - \frac{\sigma_{ls}^6}{\left(\left(x+\frac{\sqrt{2}}{2}a\right)^2+a^2\right)^3}\right] \right\} \quad (6)$$

Where R is the liquid layer radius, x is the distance between liquid and solid surface, $\varepsilon$ and $\sigma$ are energy and distance parameters in LJ 12/6 potential.

b. Inter layer energy for liquid

The cohesion force is divided into the inner layer part and inter layer part as previous. Unlike liquid layer on absolutely smooth surface, in which the self-energy-minimization determines the equilibrium distance between liquid particles, the spatial distribution here would inherit the lattice structure of solid surface, and by summing the energy between single liquid particle and those within 3 distance parameters then summing for entire liquid layer give the following equation for lattice surface.

$$W_{LL}(lattice) = \frac{9}{4}\frac{R^2\sigma_{ls}^2}{\sigma_{ll}^4}\varepsilon_{ll}\left\{4\left[\frac{\sigma_{ls}^{12}}{(x^2+0.5a^2)^6} - \frac{\sigma_{ls}^6}{(x^2+0.5a^2)^3}\right] + 8\left[\frac{\sigma_{ls}^{12}}{(x^2+2.5a^2)^6} - \frac{\sigma_{ls}^6}{(x^2+2.5a^2)^3}\right]\right\} \quad (7)$$

c. Inner layer energy for liquid

Likewise, the inner layer energy of scattered liquid layer can also be derived by summing the interaction energy between particle pairs within 3 lattice constants, and summing over the whole layer, the results derived here also needed to time 0.5 due to the mutual calculation.

$$W_{in}(lattice) = \frac{9}{4}\frac{R^2\sigma_{ls}^2}{\sigma_{ll}^4}\varepsilon_{ll}\left\{4\left[\frac{\sigma_{ll}^{12}}{a^{12}} - \frac{\sigma_{ll}^6}{a^6}\right] + 4\left[\frac{\sigma_{ll}^{12}}{2^6a^{12}} - \frac{\sigma_{ll}^6}{2^3a^6}\right] + 4\left[\frac{\sigma_{ll}^{12}}{(2a)^{12}} - \frac{\sigma_{ll}^6}{(2a)^6}\right] + 4\left[\frac{\sigma_{ll}^{12}}{5^6a^{12}} - \frac{\sigma_{ll}^6}{5^3a^6}\right]\right\} \quad (8)$$

d. supplement

As pointed out in 3.2, the inner layer structure of most liquid layers on lattice structure surface shows both distribution features, which indicates either the continuous form of energy equation described in 3.1 or the scatter form is appropriate, instead a proper combination of both equation forms should be adopted. The structural number defined previously consequently can serve as a measurement of each energy proportion. By inputting the simulation and distance parameter can acquire a preliminary energy result of a target layer, and can get a corresponding structural number. Using this initial structural number as the weight for scatter energy equation can calculate a new energy ratio for the target layer. After several times of this iteration, the structural number of the specific liquid layer would converge to $w_i$, and would serve as the weight coefficient in subsequent energy calculation. For each part of energy, the liquid layer on lattice surface should be calculated as follow:

$$W_x = w_i W_{x(lattice)} + (1 - w_i)W_{x(absolute)} \quad (9)$$

Where, subscript $x$ stands for different energy part like LL, LS etc., lattice stands for scatter form equation and absolute stands for continuous form, $w_i$ is the structural number for the layer interested.

Based on the aforementioned calculation equations can get the energy ratio data for different combinations of simulation parameters which can simulate droplets that vary from super hydrophobic to complete wetting ones, and by adopting the simulation methods described previously can get the contact angles data to represent the wettability. As described in section 3.1, similar procedures are employed here, the simulation contact angles and the corresponding energy ratios are calculated, contact angles show exponential decay tendency with energy ratio increasing. Further, the two extreme conditions where energy ratio approach 0 and 1 should be considered, and gives the following equation:

$$Contact\ angle = 180\exp(energy\ ratio/0.28) \quad (10)$$

As shown in Fig. 7, 38 different parameter combinations of simulation results on absolutely smooth surface and 27 results on lattice structure surfaces. Purple points represent the absolutely smooth simulation results v.s. energy ratio calculated, and purple line is the linear fitting results of. Green squares are lattice surfaces simulation results v.s. energy ratio calculated, and the green line is the curve exponential equation. The blue stars are the experimental results of water contact angles[23] on different metal v.s. energy ratio calculated.

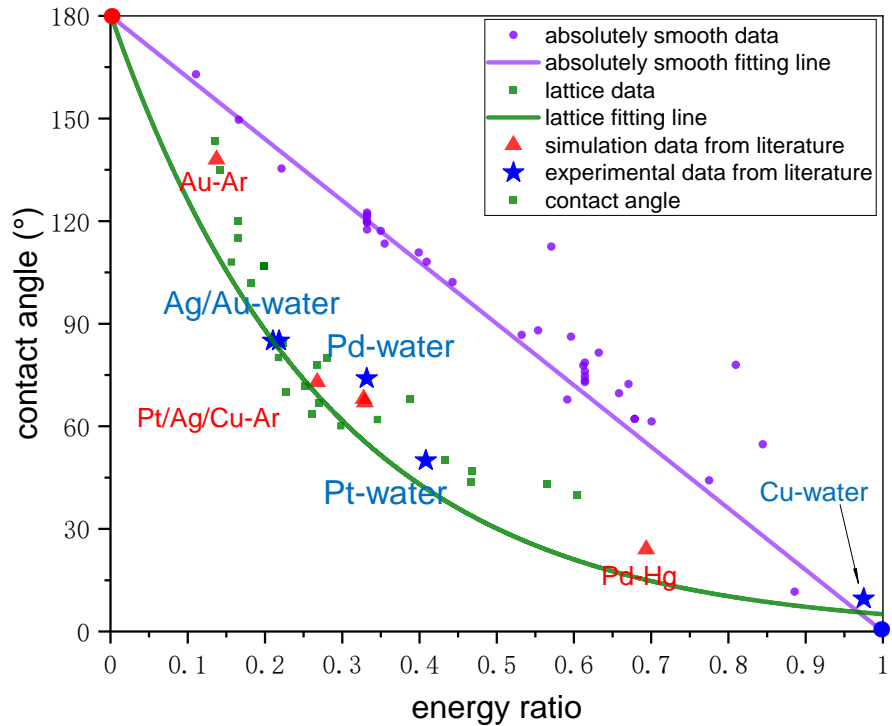

Figure 7: schematic for contact angle v.s. energy ratio; the P4 particle in MARTINI[24] force field is adopted to calculate water molecule in the comparison with experimental results, metal parameters come from literatures[25-27] and the red triangles are the simulation results in these researches.

The relationship between the competition of adhesion and cohesion forces and the wettability is obvious in Fig. 7. When energy ratio is relatively small, cohesion force of liquid particles dominates, and the liquid particles tend to stick with each other and the macroscopic performance is a larger contact angle; on the other hand when energy ratio become larger, the adhesion force gradually prevails and the liquid particles would be in favor of larger contact area with solid surface with contact angle decreasing.

For absolutely smooth surface which ignore any specific surface characteristics, the relationship between adhesion-cohesion competition and wettability is as simple as a single line with two extreme situations intercepted with two axes as expressed by eq. 5. When energy ratio is extremely small approaching zero, this means nearly no affinity exists between the solid and the liquid particles, the liquid and solid are absolutely heterogeneous and contact angle is 180 degrees. On the contrary, when energy ratio is extremely large approaching 1, the interaction between solid surface and liquid particles is nearly identical with the interaction among liquid itself, the affinity between these two materials achieve maximum and contact angle approaches 0 degree.

For surfaces with lattice structure, the two extremes remain unchanged like the absolutely smooth surfaces, but the introduction of lattice structure lead complexity into discussion. Since the lattice structure would extend to different degree into the droplet layers depending on the energy ratio values, the relation between energy ratio and contact angle has subtle deformation, and in this situation the simple linear relation deforms into exponential form. When affinity between two phases is small, the contact angle would still approach 180°. When affinity grows, even though the liquid tend to spread on solid surface, the lattice structure introduce energy barrier parallel to the surface which impedes further spreading of the liquid particles, which explains even though energy ratio continues increasing the contact angle on lattice structure surfaces will approach to 0 but won't decrease to 0 while the absolutely smooth surface exist no parallel energy barrier means the liquid can spread with no limits and the extreme contact angle is 0. Further, for surfaces with much more complicated topology and chemical composition, prediction can be made that the fundamental form of the relationship between the adhesion-cohesion competition and wettability would stay unchanged and the two extreme situations may also remain.

The relative relationship between the energy ratio which express the competition of adhesion and cohesion force and wettability could be universal, but the specific form of the relation may vary due to the specific interfacial properties. The value of the energy ratio could be derived via various methods, like quantum chemistry or density function theory to get more precise results, and by substituting into eq. 5 and 9 to get a prediction for contact angle. The energy ratio calculation for Lennard Jones particles have been proposed by current work, and such method can predict the wettability of rare gas liquid or quasi-spherical molecules.

4. conclusions

The current work employed absolutely smooth surface and lattice structure surface to distinguish the relationship between intrinsic wettability of different combinations of liquid and solid particles and the microscopic interparticle actions. The adopting of absolute smooth surface facilitates the eliminating of any specific surface characteristics and consequently give an insight into the determination of intrinsic wettability on the foundation of the express of competition between liquid-liquid cohesion force and liquid-solid adhesion force. Further the introduction of lattice structure reveals the possibility of involving various interfacial properties, and indicate that the complex characteristics of realistic surfaces (either topology or chemical composition) may introduce some kind of deviation to the linear relationship between solid – fluid energy ratio and contact angle. The prediction potential of the method introduced in present work need more confirmation, but it does provide a possible Young's-equation-free solution to predict contact angle.